\documentclass[
    ,final            
  ]
  {aipproc}

\layoutstyle{6x9}

\newcommand{\npl}{Department of Physics, University of Washington,
    Seattle, WA  98195-4290}
\newcommand{\sdt}{\sqrt{S_{\Delta T}}}
\newcommand{\rtHz}{\sqrt{\rm Hz}}
\newcommand{\ug}{(m/s$^2$/K)\,$\sdt$}
\newcommand{\tg}{{\rm Nm/K}}

\begin{document}

\title{Outgassing, Temperature Gradients and the Radiometer Effect
    in {LISA}: A Torsion Pendulum Investigation}


\author{S E Pollack}{address={\npl}}
\author{S Schlamminger}{}
\author{J H Gundlach}{}

\begin{abstract}
Thermal modeling of the LISA gravitational reference sensor (GRS)
includes such effects as outgassing
from the proof mass and its housing and the radiometer effect.
Experimental data 
in conditions emulating the LISA GRS
are required to confidently predict the GRS performance.
Outgassing and the radiometer effect are similar
in characteristics and are difficult to decouple experimentally.

The design of our torsion balance 
allows us to investigate 
differential radiation pressure, 
the radiometer effect,
and outgassing on closely separated conducting surfaces with high
sensitivity.  
A thermally controlled split copper plate 
is brought near a freely hanging plate-torsion pendulum.
We have varied the temperature on each half
of the copper plate and have measured the
resulting forces on the pendulum.

We have determined that to first order the current GRS model
for the radiometer effect, outgassing, and radiation pressure are
mostly consistent with our torsion balance measurements
and therefore these thermal effects do 
not appear to be a large hindrance to the LISA noise budget.  However,
there remain discrepancies between the predicted dependence
of these effects on the temperature of our apparatus.
\end{abstract}

\maketitle


\section{Introduction}

The low frequency end of the LISA gravitational wave sensitivity
requires that the proof masses are kept in free-fall below
$\delta a = 3 \times 10^{-15}\,({\rm m/s}^2)/\rtHz$ from
0.1 to 10\,mHz \cite{PPA, FTR, Stebbins}.  
The spurious accelerations linked
to random thermal fluctuations have been allocated one-third
of this total budget for meeting the desired LISA sensitivity \cite{Hyde}.
Thermal fluctuations on the electrode housing of the GRS 
are estimated to be
$S_{\Delta T} = 10^{-10}\,{\rm K^2}/{\rm Hz}$ \cite{FTR},
which combined with
theoretical understanding of the 
thermal effects yields
$\delta a_T = 3.5 \times 10^{-16}\,({\rm m/s}^2)/\rtHz$,
which is one-third of the allocation.
The margin in the budget that this assessment creates 
could be used to relax other requirements.
Therefore it is imperative that
the models which were used to generate the current assessment are
validated in a laboratory setting.

The three largest thermal effects in the LISA noise budget are
radiation pressure, the radiometer effect, and asymmetric outgassing.
These three terms add coherently: \cite{DRS_ITAT, Carbone}
\begin{equation}\label{eqn:accel}
\delta a_{\rm T} = \left(
\frac{8 \sigma}{3 M c} A T^3 +
\alpha_P \frac{A}{2 M}\frac{P}{T} + 
\frac{A G Q_0}{4 M} \frac{\Theta}{T^2}
\right) \sdt,
\end{equation}
\noindent 
where $\sigma$ is the Stefan-Boltzmann constant, $c$ is the speed of light,
$M = 1.96\,$kg is the mass of the LISA proof mass, 
$A = 2.12 \times 10^{-3}\,{\rm m}^2$ is the area of one face of the LISA proof mass,
$P = 10^{-5}\,$Pa is the residual gas pressure at the proof mass, and
$T = 293\,$K is the temperature. 
The first term in Equation~\ref{eqn:accel} 
is due to fluctuating asymmetric radiation
pressure, $\delta a_{\rm RP}$.
The second term is due to the radiometer effect, $\delta a_{\rm RE}$.  
The pressures
on opposite faces of the proof mass equilibrate quickly so 
that there is nearly no pressure difference across it.
This implies that the contribution due to the radiometer effect
should be reduced \cite{Ruediger}.  
Leading to a 
pressure reduction factor of 
$\alpha_P \approx 0.1$ \cite{Dolesi}.
The last term is due to fluctuations in asymmetric outgassing
on opposite sides of the proof mass.
The outgassing rate $Q (T)$ most likely will follow an exponential law in
temperature referenced to an activation temperature $\Theta$.
The value of this activation temperature should be no more than
$30 000$\,K \cite{Stebbins}.  The value for the nominal outgassing
rate is taken to be 
$Q_0 \approx 4 \times 10^{-10}\, {\rm Torr\ L\ cm}^{-2}\ {\rm s}^{-1} 
     \approx 5 \times 10^{-7}\,{\rm kg/s}^3 $ \cite{DRS_ITAT}.
The factor, $G$, 
is related to gas particle
conductances about the proof mass and through holes in the
housing walls, 
and to the outgassing area of the proof mass.
The nominal value for LISA is
$G = 0.4\,$m/s \cite{DRS_ITAT}.

The derivations of these three effects were made assuming
a constant temperature gradient across the housing which is then
varied.  This implies that the three effects described above
are dependent on the frequency of fluctuations 
only through the level of the thermal fluctuations themselves, $\sdt$.


Our apparatus 
provides the unique opportunity
to investigate the gap spacing dependence as well as the pressure
and linearity of the thermal effects.
Our results are complimentary to those that have been
presented by the University of Trento group \cite{Carbone, Mauro}.  
Whereas they have designed a mock-up of the LISA proof mass and electrode
housing and therefore emulate the geometry of LISA, 
we have an increased torque sensitivity, the ability
to change the gap spacing, and a simpler geometry which allows
ease in modeling \cite{Stephan}.

\section{Torsion Balance Apparatus}

Our torsion balance apparatus is described in \cite{Stephan}.
Each of the two halves of the copper plate
has embedded a heating element and a temperature sensor.
Further temperature sensors are located
on the plate translation stage, 
    near the control electrodes, on the autocollimator,
on the outside of our vacuum chamber,
on the inside of the styrofoam enclosure, and the
temperature controlled laboratory.


We predict values for the three thermal noise sources 
under study for our torsion balance tests.
Since our torsion pendulum measures torques we must
convert the measured torques into LISA-equivalent accelerations.
As mentioned in \cite{Stephan}, this conversion is done by
using
a moment arm at which the spurious forces act on the pendulum and
the mass of one LISA proof mass.
The base pressure of our experiment is
$\approx1.2 \times 10^{-5}\,$Pa.
We assume that the outgassing activation
temperature
and the outgassing rate of copper is similar to that assumed for the
LISA GRS.
The cross-sectional area overlap of one half of the copper plate
and our pendulum is 
$A_{\rm UW} = 4.7 \times 10^{-3}\,\rm{m}^2$.
The gas conductance rate for our setup should be similar in value
to those for the LISA GRS, likewise we assume
the same value for the factor $G$.
Note that simple gas dynamics implies that the conductance
rate should increase if the separation between the pendulum and plate
grows, and therefore outgassing effects should decrease with plate-pendulum
separation.

In Table~\ref{tab:exp} we list the predicted values for the three
thermal effects in Equation~\eqref{eqn:accel} for LISA 
and the predicted values for our experiment.
In this table we have computed our LISA-equivalent noise levels
when the mean temperature of the copper plate in our apparatus
is 298\,K and when it is 301\,K.
From these values we predict a thermally induced
acceleration noise of $3 \times 10^{-11}$ \ug\ relatively
independent of the temperature of the copper plate.
In addition, we show values for the 
system at an elevated pressure of $1.6 \times 10^{-4}\,$Pa,
for which we expect an 
increase by nearly a factor of 2.

\begin{table}
\begin{tabular}{lcccccc}
\tablehead{1}{l}{b}{Noise Source} & &
\tablehead{1}{c}{b}{LISA} &
\tablehead{1}{c}{b}{UW (25$^\circ$C)} &
\tablehead{1}{c}{b}{UW (28$^\circ$C)} &
\tablehead{1}{c}{b}{UW (25$^\circ$C)} &
\tablehead{1}{c}{b}{UW (28$^\circ$C)}\\
& & & \multicolumn{2}{c}{low pressure} & \multicolumn{2}{c}{high pressure}\\
	\hline
Radiation Pressure & $\delta a_{\rm RP}$ & 1.4 & 1.5 & 1.6 & 1.5 & 1.6\\
Radiometer Effect  & $\delta a_{\rm RE}$ & 0.2 & 0.2 & 0.2 & 2.8 & 2.8\\
Outgassing	   & $\delta a_{\rm OG}$ & 1.4 & 1.3 & 1.2 & 1.3 & 1.2\\
       \hline
Coherent Total	   & $\delta a_{\rm T}$  & 3.0 & 3.0 & 3.0 & 5.6 & 5.6\\
\end{tabular}
\caption{ {\bf Expected Values of Thermal Noise Sources} for LISA
using the values given in the introduction, and 
LISA-equivalent accelerations for our apparatus
using the values given in the text.
The mean temperature of the copper plate and the 
residual gas pressure differentiates each column.
The low and high pressure data were computed with 
base pressures of $1.2 \times 10^{-5}\,$Pa and 
$1.6 \times 10^{-4}$\,Pa respectively.
Values in the table are in units of $10^{-11}$\,\ug. }
\label{tab:exp}
\end{table}

We present our results in terms of a 
thermal-to-torque transfer function.  
The conversion from torque to LISA-equivalent acceleration is 
$\approx 15\,({\rm m/s}^2)/({\rm N m})$, 
such that the expected thermal-to-torque
transfer function should be $2 \times 10^{-12}$ \tg\ 
when the system is at low pressure, 
and about $4 \times 10^{-12}$ \tg\ for the high pressure scenario.  

The copper plate reached 35$^\circ$ during
the two day mild bakeout of our apparatus under vacuum.
Our measurements were made 150 days after this bakeout occurred.
The copper plate is made of oxygen-free high-conductivity (OFHC) copper,
which has an outgassing rate 10 times lower than that
assumed in the above calculations \cite{Perkins}.

\section{Thermal Measurements}

The temperature difference between the two halves of
the copper plate emulates a
\emph{temperature gradient} across the LISA proof mass.
Our procedure is to vary this gradient
recording the torque,
thereby the LISA-equivalent acceleration,
as a function of a variety of adjustable variables,
such as the temperature gradient, the mean temperature of the copper plate, 
the plate-pendulum separation, and the residual gas pressure inside
our apparatus.

\subsection{DC Measurements}

\begin{figure}
  \includegraphics[width=.5\textwidth]{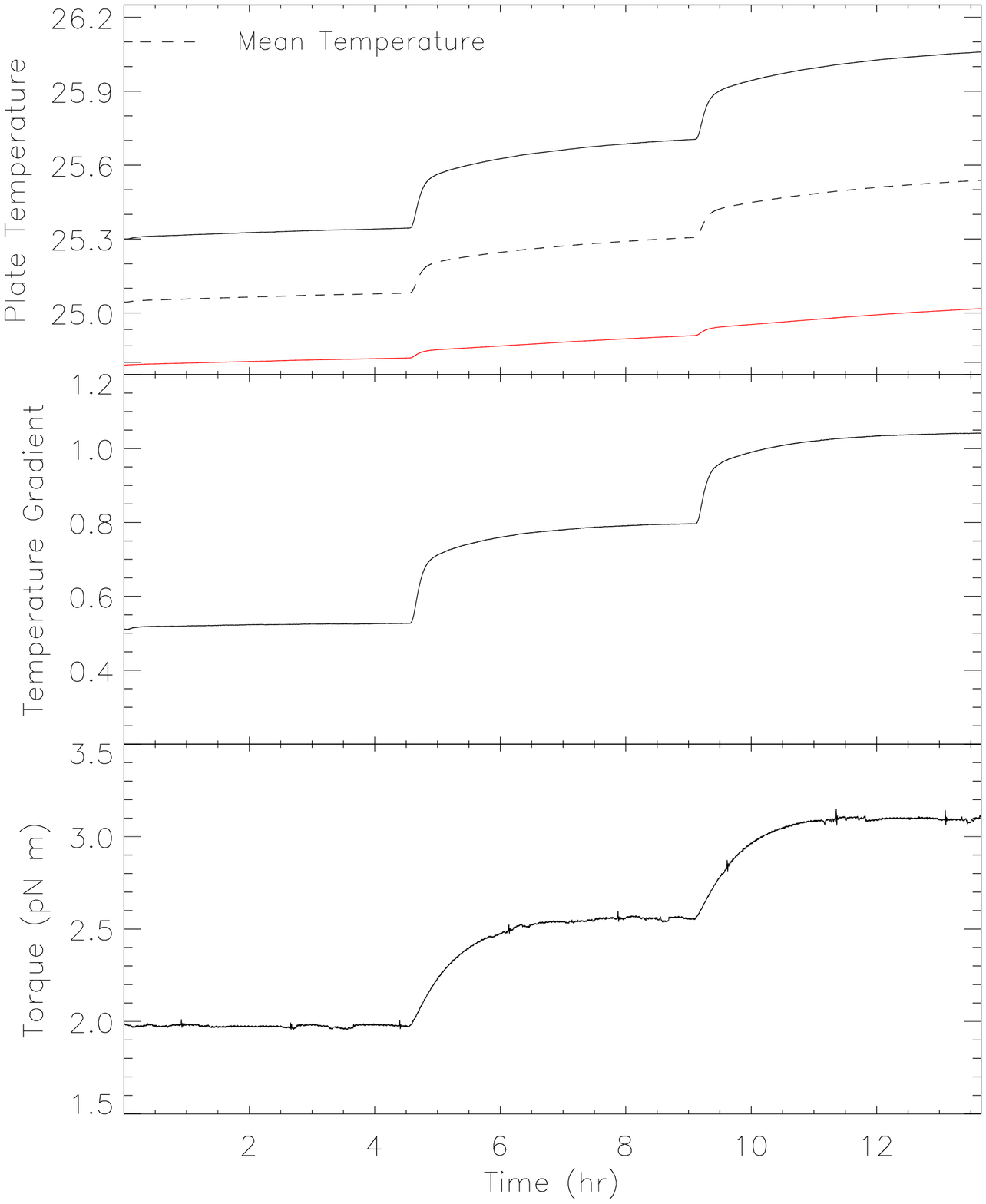}
  \includegraphics[width=.5\textwidth]{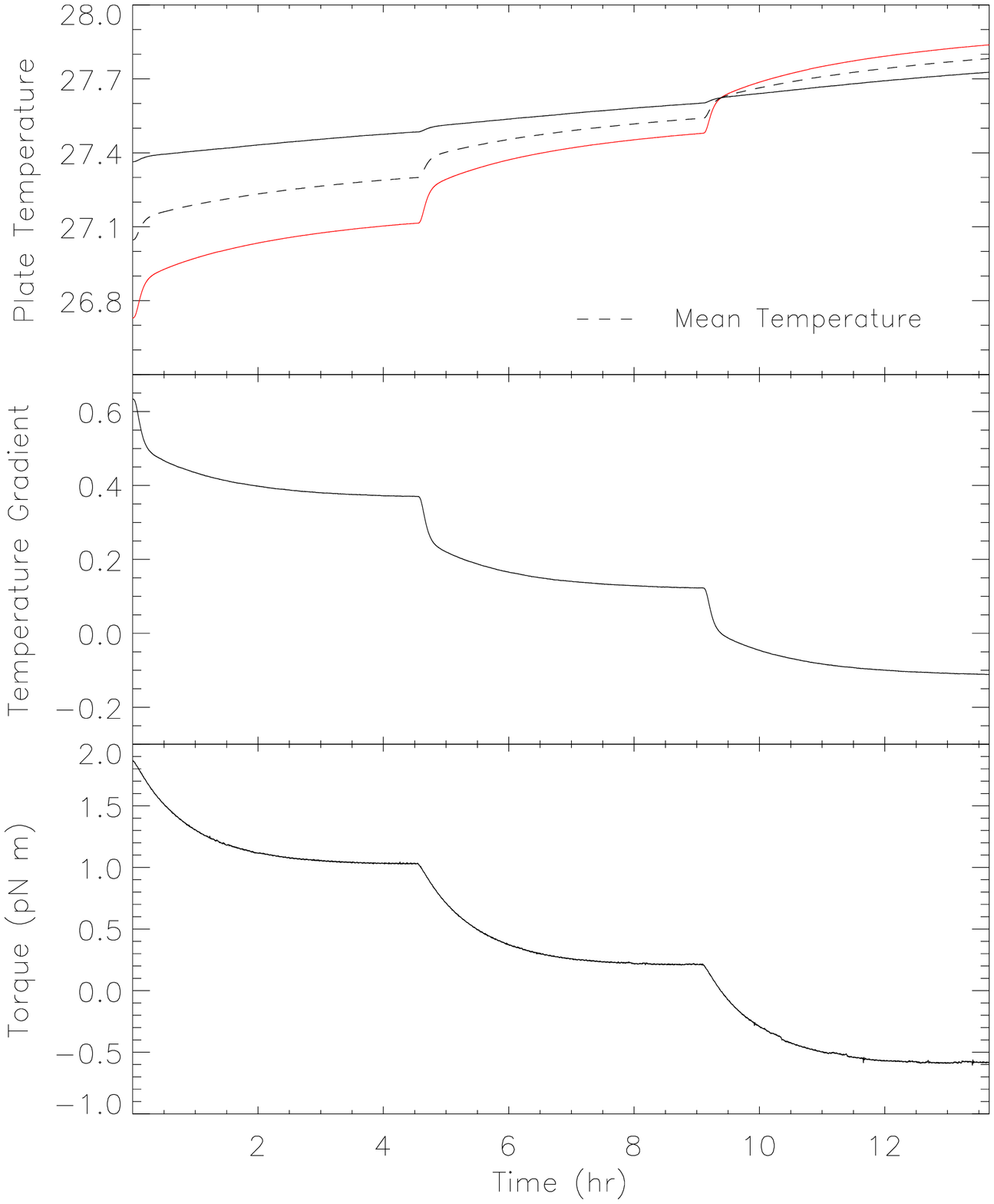}
\caption{Heating of one half of the copper plate (left) 
followed by heating of the other half some time later (right)
in equals steps of heater output power.
The top chart shows the temperature of each half of the
copper plate as well as the mean temperature all in $^\circ$C,
the middle chart shows the temperature gradient between the
two halves in K, and the lower chart shows the measured torque
on the pendulum.  The plate-pendulum separation
for this data was 1\,mm.
}\label{fig:dc}
\end{figure}

Applying constant power to the heater in 
either the left or right section
of the copper plate
shows a linear rise in temperature of that half as
well as a linear response of the torque on the pendulum.
Figure~\ref{fig:dc} shows heating of half of the copper plate in
constant steps of heater power.
This data was taken with a plate-pendulum separation of 1\,mm.
The heated half changes by about 0.76\,K over the course
of this data set.
Thermal coupling
from one half of the copper plate to 
the other is $\sim\!\!30$\% such that the other half gained 0.23\,K.
The temperature sensor placed on the electrode side of the
apparatus (inside the vacuum chamber)
rose by
47\,mK
over the course of this data set.
Temperature sensors
attached to the outside of the vacuum chamber changed in temperature
by $\sim\!\!46$\,mK.  The temperature of the styrofoam enclosure
did not change appreciably.  This indicates that the thermal feedthrough
from the copper plate to the apparatus is on the order of 10\%.

It can be seen from the slower rise in the torque 
as compared with the temperature gradient
that there is a delay between
the temperature gradient and the reaction of the feedback loop
of $\sim\!\!1700$\,s.
We suspect that this delay may be due to
an effective activation time of outgassing: the trapped gas at the
surface of the copper immediately escapes whereas the trapped gas within
the copper slowly diffuses.  
Additionally, 
transfer of heat to the pendulum
should raise the pendulum temperature by a fraction of the plate
temperature change.  The heating of the pendulum should increase
outgassing from the silicon.  
There is a delay between the heating of the copper plate and the heating of the
pendulum 
leading to a delayed
response in the feedback torque.  

The temperature gradient between the two halves of the split copper
plate is the important quantity under investigation.
The change in temperature gradient 
makes a change in all three thermal sources under investigation.
This causes a pronounced change in the torque.  The thermal-to-torque
transfer function for the 
data on the left in Figure~\ref{fig:dc} is about 2.3~p\tg.

After some time we heated the other half of the copper plate and brought the
temperature gradient back to nearly zero.  This data
is on the right in Figure~\ref{fig:dc}.  
The thermal-to-torque transfer function for this
data is 3.0~p\tg.
Notice that the mean temperature of the copper plate in the 
data on right is $\sim\!\!1.5$\,K higher than the data on the left.
This increase in the transfer function 
appears to be a function of the mean
temperature of the copper plate as described in more detail below.

\subsection{AC Measurements}

\begin{figure}
  \includegraphics[height=.5\textheight, angle=90]{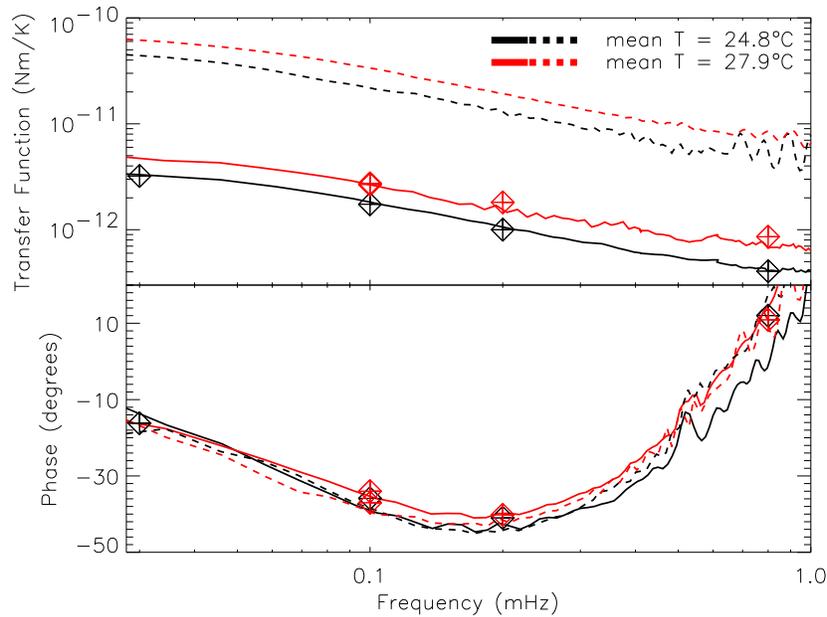}
\caption{Thermal-to-torque transfer function and phase delay
for thermal frequency sweeps between $30\,\mu{\rm Hz}$ and 1\,mHz.
The solid (dashed) transfer function traces were taken
at a base pressure of $1.2 \times 10^{-5}\,$Pa ($1.6 \times 10^{-4}$\,Pa).
Raising the gas pressure increased
the transfer function by a factor of 13.
Raising the mean temperature of the copper plate by 3\,K raised the
transfer function by a factor of $1.5$ irregardless of the gas
pressure in our apparatus.  
} \label{fig:ac}
\end{figure}

The important frequency band of thermal measurements for LISA is
from 0.1 to 1\,mHz.  Looking at coherent signals allows us to determine
the thermal frequency response of our system.  By sweeping
in frequency from 0.1 to 1\,mHz we can 
map
the transfer function from thermal input to torque output.
Figure~\ref{fig:ac} contains data for such a frequency sweep
at two mean plate temperatures and two gas pressures.
This data was taken with a plate-pendulum separation of 1\,mm.

The thermal mass of the copper plate creates a low pass filter
so that the temperature gradient amplitude is smaller at
higher frequencies.
Our typical amplitude is nearly 200\,mK at 0.1\,mHz.
From a series of measurements we found that the thermal-to-torque 
response is independent of the signal amplitude at any particular
frequency \cite{thermals}.  
However, we still observe an apparent 
low pass response with a corner
frequency near 0.1\,mHz and a slope index of $-0.75$ in amplitude.

From our DC measurements we found a time delay between the temperature
gradient and the control torque of about 1700\,s.  At 10\,$\mu$Hz
this would be a phase delay of $-18^\circ$ which is in good agreement
with the data presented in Figure~\ref{fig:ac}.
This phase delay becomes larger for frequencies
above 1\,mHz following the decrease in transfer function.  
However, it appears to turn around at higher frequencies.
We suspect this is an experimental artifact due to electrical
coupling in our system.

To verify that this frequency and phase dependence was not
dependent on our frequency sweeping function, we used a variety of different
sweep functions, such as linear, logarithmic, and more complicated
functions of time, as well as monochromatic sinusoids.
We found no dependence on the sweep function.
The individual data points 
in Figure~\ref{fig:ac} are for fixed-frequency sine waves.

\section{Discussion}


When the mean temperature of the copper plate is elevated
by a few degrees
we see an increase in the thermal-to-torque transfer function,
which is not predicted by Table~\ref{tab:exp}.
An increase in the outgassing rate is likely the cause.
As mentioned above, it is also possible that the pendulum is
appreciably changing in temperature such that the outgassing
rate from the silicon increases as well.  
The outgassing rate should decay with time 
as a power law \cite{Perkins, outgas}.
Observing this decay would yield support for this hypothesis.



The radiometer effect is linearly proportional to the
gas pressure in the system.  

Figure~\ref{fig:ac} contains thermal-to-torque traces at 
an elevated pressure of $1.6 \times 10^{-4}$\,Pa.
The thermal-to-torque transfer function 
at frequencies above 0.1\,mHz appears to be a factor
of 13 higher when the system is at this higher pressure.
This linear response is exactly expected for the radiometer effect,
if the radiometer effect accounted for 
the bulk of the thermal-to-torque transfer.
As demonstrated in Table~\ref{tab:exp}, the radiometer effect should
be a small contribution to the total.
The three contributing thermal effects add linearly since they
are coherent in the temperature gradient.
If outgassing and radiation pressure are independent of gas pressure,
then the increase in the transfer function when the system is at high
pressure would be due to the radiometer effect.  This
would imply that the radiometer effect is
accounting for the bulk of the transfer,
and therefore is a factor of 15 higher than
anticipated, i.e., the pressure reduction factor, $\alpha_P$, should be an 
amplification of 1.5 not a reduction of 10.  
However, the radiometer effect is inversely proportional to 
the mean temperature and therefore
does not account for the increase of $1.5$ when the copper
plate is at an elevated temperature.




If instead outgassing is the cause of the increase when the system is warm,
then the outgassing contribution must also increase when the system
is at an elevated pressure which is currently
not included in the LISA models \cite{DRS_ITAT}.

\section{Directions for Further Study}

Our results are reasonably consistent with the 
current theoretical understanding of the thermal effects
used in modeling for the LISA GRS.  However, there does remain
an anomalous increase in the thermal-to-torque transfer function
when the copper plate is heated by a small amount regardless
of base pressure.  
We are currently conducting measurements to help explain
this.
In addition, we are in the process of making
measurements over a range of plate-pendulum separations
to determine the distance dependence of the thermal effects.

\begin{theacknowledgments}
We would like to thank the members of E\"{o}t-Wash
and the Center for Experimental Nuclear Physics and Astrophysics
at the University of Washington
for infrastructure.
This work has been performed under contracts 
NAS5-03075 through GSFC and 
1275177 through JPL,
and through NASA Beyond Einstein grant NNG05GF74G.
\end{theacknowledgments}

\nocite{*}
\bibliographystyle{aipproc}   

\bibliography{lisa6}

\end{document}